\documentclass[aps,prd,twocolumn,superscriptaddress,preprintnumbers,floatfix,nofootinbib,notitlepage,showkeys,showpacs]{revtex4-1}

\usepackage[utf8]{inputenc}

\usepackage{graphicx}
\usepackage{latexsym}
\usepackage{amsmath}
\usepackage{amssymb}
\usepackage{bbm}
\usepackage{ulem}
\usepackage{pdfsync}
\usepackage{epsfig}
\usepackage{epstopdf}
\usepackage{subfigure}
\usepackage{color}
\usepackage{comment}
\usepackage{slashed}
\usepackage{multirow}

\begin{document}

\title{Kinetic Sunyaev-Zel'dovich tomography with line-intensity mapping}

\author{Gabriela Sato-Polito}
\email{gsatopo1@jhu.edu}
\affiliation{Department of Physics and Astronomy, Johns Hopkins University, \\
Baltimore, MD 21218, USA}

\author{José Luis Bernal}
\affiliation{Department of Physics and Astronomy, Johns Hopkins University, \\
Baltimore, MD 21218, USA}

\author{Kimberly K. Boddy}
\affiliation{Theory Group, Department of Physics, The University of Texas at Austin, Austin, TX 78712, USA}

\author{Marc Kamionkowski}
\affiliation{Department of Physics and Astronomy, Johns Hopkins University, \\
Baltimore, MD 21218, USA}

\begin{abstract}
The kinetic Sunyaev-Zel'dovich (kSZ) effect is a secondary cosmic microwave background (CMB) anisotropy induced by the scattering of CMB photons off intervening electrons. Through cross-correlations with tracers of large-scale structure, the kSZ effect can be used to reconstruct the 3-dimensional radial-velocity field, a technique known as kSZ tomography. We explore the cross-correlation between the CMB and line-intensity fluctuations to retrieve the late-time kSZ signal across a wide redshift range. We focus on the CII emission line, and predict the signal-to-noise ratio of the kSZ tomography signal between redshifts $z=1-5$ for upcoming experiments. We show that while instruments currently under construction may reach a low-significance detection of kSZ tomography, next-generation experiments will achieve greater sensitivity, with a detection significance of $\mathcal{O}(10^2-10^3)$. Due to sample-variance cancellation, the cross-correlation between the reconstructed velocity field from kSZ tomography and intensity fluctuations can improve measurements of 
contributions from new physics to the power spectrum at large scales. To illustrate this improvement, we consider models of the early Universe that induce primordial local-type non-gaussianity and correlated compensated isocurvature perturbations. We show that with CMB-S4 and an AtLAST-like survey, the uncertainty on $f_{\rm NL}$ and $A_{\rm CIP}$ can be reduced by a factor of $\sim 3$, achieving $\sigma(f_{\rm NL}) \lesssim 1$. We further show that probing both low and high redshifts is crucial to break the degeneracy between the two parameters.
\end{abstract}

\preprint{UTTG-21-2020}

\maketitle

\section{Introduction}
The cosmic microwave background (CMB) anisotropies have been measured with increasing precision and are foundational to our current understanding of the standard cosmological model. On large scales, the CMB temperature power spectrum is dominated by primary anisotropies\textemdash those produced at the surface of last scattering\textemdash and the temperature fluctuations have been measured close to their ultimate cosmic variance limit \cite{Aghanim:2018eyx}. However, on smaller angular scales, the leading contributions to the CMB power spectrum are secondary anisotropies, which are produced as CMB photons travel across the Universe and interact with intervening matter. These secondary anisotropies are at the forefront of current observational developments and will be measured with much higher significance with the next generation of CMB experiments \cite{Ade:2018sbj, Abazajian:2016yjj, Sehgal:2019ewc}.

The kinetic Sunyaev-Zel'dovich (kSZ) effect is a secondary anisotropy induced by the scattering of CMB photons off intervening electrons with bulk peculiar motion relative to the CMB frame \cite{1980MNRAS.190..413S}. For multipoles $\ell \gtrsim 4000$, the kSZ effect becomes the dominant black-body component of the CMB. The kSZ effect has two main contributions: patchy kSZ, produced during the epoch of reionization, and late-time kSZ, generated by scattering off electrons in ionized gas present in halos in the post-reionization Universe. The latter contribution has been detected both in the CMB temperature power spectrum \cite{George:2014oba} and by cross-correlating the CMB temperature anisotropies with galaxy surveys \cite{2012PhRvL.109d1101H, Ade:2015lza, Soergel:2016mce, DeBernardis:2016pdv, Schaan:2020qhk}.

Cross-correlations with large-scale structure are key to fully exploit the cosmological information encoded in the secondary CMB anisotropies. In particular, this cross-correlation enables the determination of the kSZ contribution as a function of redshift, a technique called kSZ tomography \cite{Ho:2009iw, Shao:2010md, Zhang:2010fa, Zhang:2000wf, Munshi:2015anr, Hill:2016dta}. Since the kSZ effect is, to lowest order, a Doppler shift, this cross-correlation can be used to reconstruct the line-of-sight velocity field of the electrons. A variety of cosmological applications for measurements of the large-scale velocity field have been proposed, such as constraining primordial non-Gaussianity~\cite{Munchmeyer:2018eey}, testing modified gravity \cite{Xu:2014doa, Mueller:2014nsa, Bianchini:2015iaa}, probing the properties of dark energy and dark matter \cite{Xu:2013jma, DeDeo:2005yr, HernandezMonteagudo:2005ys, Bhattacharya:2007sk}, measuring neutrino masses \cite{Mueller:2014dba}, testing homogeneity \cite{Zhang:2015uta, 2011PhRvL.107d1301Z}, etc. The velocity field is an unbiased tracer of the matter density and can therefore also be an important additional probe for multi-tracer cosmological analyses.

Galaxies are currently the sole tracers of large-scale structure employed for a detection of the kSZ effect. However, this situation is expected to change with the recent developments in line-intensity mapping (LIM). LIM is a technique that measures the integrated emission of atomic or molecular spectral lines along the line-of-sight. While galaxy surveys become less effective at high redshifts as sources become too faint to be resolved, LIM is sensitive to all emitters, whether they can be individually resolved or not. Since photons emitted at larger distances are more highly redshifted, it enables the construction of three-dimensional maps of the Universe.

Intensity mapping has the potential to deliver statistical measurements of large-scale structure at high redshifts. In cross-correlation with the CMB, this technique is uniquely suited to probe the velocity field deep into the observable Universe. Previous works have focused on cross-correlating the CMB with the 21-cm neutral hydrogen line to measure the patchy kSZ signal \cite{Ma:2017gey, Roy:2019qsl, LaPlante:2020nxx} and on mitigating foregrounds and other systematic effects in order to retrieve the kSZ signal at low redshifts \cite{Li:2018izh}.

In this work, we investigate the potential to measure late-time kSZ tomography using the cross-correlation between LIM and the CMB across the full post-reionization Universe. To demonstrate this technique, we focus on intensity-mapping experiments that target the fine-structure line in singly ionized carbon (CII), with a rest-frame frequency of $\nu=1901.0$~GHz. We consider LIM and CMB experiments on two timescales: those currently under construction, with first light expected in the next few years, and next-generation experiments expected to be operating in the next decade. The near-future LIM and CMB experiments we consider are the Epoch of Reionization Spectrometer (EoR-Spec) on CCAT-prime \cite{2020JLTP..199.1089C} and Simons Observatory (SO) \cite{Ade:2018sbj}. For the next-generation experiments, we consider a survey design inspired by AtLAST \cite{2019BAAS...51g..58K} and CMB-S4 \cite{Abazajian:2016yjj}.

While a kSZ detection is beyond reach for the first-light CCAT-prime design, we show that an upgraded ``Phase II'' design can achieve a $3\sigma$ detection of the kSZ signal in cross-correlation with SO at the redshift of $z\sim 3.7$. We expect substantial improvements from next-generation instruments, yielding a detection significance of $\mathcal{O}(10^2-10^3)$.

To illustrate the benefits of measuring kSZ tomography with LIM, we assess its potential to test theories of the early Universe. Primordial non-Gaussianity (PNG) of the local type and correlated compensated isocurvature perturbations (CIPs) are two potential signatures of inflation. Both induce a contribution to the halo bias with the same scale dependence, but with different redshift evolution~\cite{Dalal:2007cu, Barreira:2019qdl}. Reference~\cite{Barreira:2020lva} showed that this degeneracy can be broken by considering galaxy samples with different biases, while Ref.~\cite{Hotinli:2019wdp} showed that using kSZ tomography in correlation with galaxies can capture this redshift evolution. Here we show that this degeneracy can be more efficiently broken by probing a wider redshift range, enabled by using intensity mapping instead of galaxies to perform the kSZ tomography.

We forecast the uncertainties on the measurements of the contributions of PNG and CIPs to the halo bias with and without the reconstructed velocity field. For the fiducial instrument design, we find that the uncertainty on the determination of both contributions can be improved by a factor of $3$ by the addition of the reconstructed velocity field with LIM. When only PNG is considered, the observational goal of $\sigma(f_{\rm NL}) \lesssim 1$ (see, e.g., Ref.~\cite{Alvarez:2014vva}) can be achieved. We show that including a wide range of redshifts is key to breaking the degeneracy between PNG and CIPs, achievable with intensity mapping. While measurements at high redshifts ($z \gtrsim 1.5$) can offer tighter constraints on the relevant parameters, a combination with low redshifts ($z \lesssim 0.75$) is essential due to their different degeneracies.

This paper is organized as follows. In Section~\ref{sec:ksz} we describe the approach adopted to model the signal and noise for kSZ tomography, and the model for the clustering of the intensity fluctuation is outlined in Section~\ref{sec:LIM}. In Section~\ref{sec:SNR} we discuss the LIM and CMB instruments considered in this work, and show the signal-to-noise ratio for kSZ tomography measurements and for the reconstructed velocity field. The effect of PNG and CIPs on the halo bias and constraints on the corresponding parameters are shown in Section~\ref{sec:PNG&CIP}. We conclude in Section~\ref{sec:conclusion}.

We adopt the standard $\Lambda$CDM cosmology as our fiducial model throughout, with the following parameters from \textit{Planck} 2018 \cite{Aghanim:2018eyx}: a reduced Hubble constant $h=0.674$, an optical depth to reionization $\tau_{\rm re} = 0.054$, matter and baryon density parameters today $\Omega_m = 0.315$ and $\Omega_b = 0.049$, respectively, and spectral index and amplitude of the primordial scalar power spectrum $n_s = 0.965$ and $A_s= 2.2\times 10^{-9}$, respectively.

\section{The kSZ Effect}\label{sec:ksz}

The temperature fluctuation induced by the kSZ effect in the $\hat{\pmb{n}}$ direction on the sky is given by
\begin{equation}
    T(\hat{\pmb{n}}) = -T_{\mathrm{CMB}} \sigma_T \int \frac{d\chi}{1+z} e^{-\tau(\chi)}  n_e(\hat{\pmb{n}}, \chi)\ \hat{\pmb{n}} \cdot \pmb{v},
\end{equation}
where $T_{\mathrm{CMB}}$ is the mean CMB temperature today, $\sigma_T$ is the Thomson scattering cross section, and $\tau$ is the optical depth from the observer to a scatterer with peculiar velocity $\pmb{v}$ located at a comoving distance $\chi$ at redshift $z$. The electron number density can be written as $n_e(\hat{\pmb{n}}, \chi) = \bar{n}_e(\chi) [1+\delta_e(\hat{\pmb{n}}, \chi)]$, where $\bar{n}_e$ is the average electron number density and $\delta_e$ is the electron density perturbation.

In order to extract redshift information, the CMB map can be cross-correlated with a tracer of the large-scale structure, in a procedure known as kSZ tomography. While several statistics have been proposed to measure the kSZ tomography, Ref.~\cite{Smith:2018bpn} showed that most of these various approaches are equivalent to a bispectrum estimation of the form $\langle \delta \delta T \rangle$. This technique can be used to reconstruct the radial velocity field, which can in turn be added to cosmological analyses as an additional matter tracer. In what follows we summarize some of the main results we use. The kSZ estimator and an outline of the derivation of the kSZ bispectrum are shown in App.~\ref{app:kSZ}, but we refer the reader to Ref.~\cite{Smith:2018bpn} for further details.

\subsection{Estimator and Noise}
The fundamental statistical quantity that carries the kSZ tomography signal is the 3-point function involving two powers of the overdensities $\delta_{\mathrm{X}}$, corresponding to a tracer X of large-scale structure, and one power of the integrated temperature fluctuation induced by the kSZ effect $T$, i.e.\ $\langle \delta_{\mathrm{X}} \delta_{\mathrm{X}} T\rangle$. It can be shown that the squeezed limit dominates the kSZ bispectrum signal, which we discuss in greater detail in App.~\ref{app:kSZ}. In this limit, the bispectrum is given by
\begin{equation}
B(k_L,k_S,\ell,k_{Lr}) = -\frac{K_* k_{Lr}}{\chi^2_*} \frac{P_{{\rm X}v}(k_L)}{k_L} P_{{{\rm Xe}}}(k_S),
\label{eq:squeezed-kSZ}
\end{equation}
where $P_{{\rm X}v}$ is the cross-correlation between $\delta_{{\rm X}}$ and the velocity field, $P_{{\rm X}e}$ is the cross-correlation with the electron density perturbations, $k_L$ is the long-wavelength mode (with $k_{Lr}$ as its component along the line of sight),  $k_S$ is the short-wavelength mode, fulfilling $k_L \ll k_S$, and $\ell$ is the spherical harmonic multipole. We have defined
\begin{equation}
    K_* \equiv -T_{{\rm CMB}} \sigma_T \bar{n}_{e,0} e^{-\tau(\chi_*)} (1+z_*)^2,
\end{equation}
where the subscript $*$ denotes quantities evaluated at redshift $z_*$, and $\bar{n}_{e,0}$ is the mean electron number density today.

The total signal-to-noise ratio $S/N$ of the kSZ bispectrum in the squeezed limit is given by
\begin{equation}
\begin{split}
\left(\frac{S}{N}\right)^2 =& V\frac{K^2_*}{8\pi^3 \chi^2_*}\left[\int dk_L \int dk_{Lr} \frac{k_{Lr}^2}{k_L} \frac{P^2_{{{\rm X}v}}(k_L, k_{Lr})}{P^{{\rm tot}}_{{{\rm XX}}}(k_L, k_{Lr})} \right] \\ &\times\left[\int dk_S \, k_S \frac{P^2_{{{\rm X}e}}(k_S, k_{Lr})}{P^{{\rm tot}}_{{{\rm XX}}}(k_S, k_{Lr})} \frac{1}{C^{{\rm tot}}_{\ell = k_S\chi_*}}\right],
\label{eq:SNR}
\end{split}
\end{equation}
where $V$ is the survey volume, $P^{\rm tot}_{\rm XX}$ is the total auto-power spectrum for the tracer X including all relevant sources of noise, $C^{\rm tot}_{\ell}$ is the total delensed CMB temperature power spectrum including instrument noise, and the integral over $k_{Lr}$ is from $-k_L$ to $k_L$. The relevant integration regions for $k_L$ and $k_S$ are $k_L \lesssim 0.1$~Mpc$^{-1}$ and $1 \lesssim k_S \lesssim 5$~Mpc$^{-1}$.

We include in the calculation of $C^{\rm tot}_{\ell}$ the contribution from primary CMB anisotropies, late-time kSZ, and instrumental noise. We compute the contribution of the kSZ effect to the CMB following the standard kSZ modeling, using the fully non-linear electron distribution (see, e.g., Refs.~\cite{Hu:1999vq, Park:2015jea} and App.~B in Ref.~\cite{Smith:2018bpn}). The instrumental noise power spectrum of the CMB experiment is modelled as
\begin{equation}
\label{eq:CMB_noise}
    N_{\ell}^{\rm CMB} = s^2 \exp \left[\frac{\ell(\ell+1)\theta_{{\rm FWHM}}^{\rm CMB}}{8\ln 2}\right],
\end{equation}
where $\theta_{{\rm FWHM}}^{\rm CMB}$ is the beam profile full width at half maximum of the CMB experiment, and $s$ is the map temperature sensitivity.

\subsection{Velocity Reconstruction}
A quadratic estimator for the large-scale modes of the velocity field can be obtained by summing over the small-scale modes of the CMB temperature and matter tracer fluctuations \cite{Deutsch:2017ybc}. This method, as shown in Ref.~\cite{Smith:2018bpn}, is equivalent to the the optimal kSZ bispectrum estimator.

The noise in the reconstruction of the radial velocity field is \cite{Smith:2018bpn}
\begin{equation}
\label{eq:Nvr}
N_{v_r}(k_{Lr}) = \frac{\chi^2_*}{K^2_*} \left[ \int \frac{dk_S\, k_S}{2\pi} \frac{P^2_{{\rm X}e}(k_S,k_{Lr}) }{P^{{\rm tot}}_{{\rm XX}}(k_S,k_{Lr})C^{\rm tot}_{\ell=k_S \chi_*}} \right]^{-1}.
\end{equation}
Notice that the reconstruction of the large-scale modes of the radial velocity field depends on the sensitivity to small-scale modes, and $k_{Lr}=k_{Sr}$ due to symmetry (see App.~\ref{app:kSZ}). The noise for the total (radial and transverse components) velocity is then given by
\begin{equation}
\label{eq:Nvv}
N_{vv}(k_L, k_{Lr}) = \left(\frac{k_{Lr}}{k_L}\right)^2 N_{v_r}(k_{Lr}).
\end{equation}

\subsection{Extracting cosmological information}
The kSZ effect can offer new insights into both astrophysics and cosmology due to the connection between electron density, its velocity field, and the CMB. As shown in Eq.~\eqref{eq:squeezed-kSZ}, kSZ tomography is sensitive to the small-scale electron distribution within dark matter halos and to the large-scale velocity field of electrons. The latter is of primary interest for cosmology.

At linear order, the matter velocity and density fields are linked by the relation
\begin{equation}
    \pmb{v}(\pmb{k}) = \hat{\pmb{k}} \frac{faH}{k} \delta(\pmb{k}),
\end{equation}
where $f$ is the growth rate, $a$ is the scale factor, and $H$ is the Hubble parameter. The velocity field is therefore an unbiased tracer of the matter density field and can be combined with a galaxy or LIM survey to perform a multi-tracer analysis. This technique can be used to achieve sample variance cancellation \cite{Seljak:2008xr}, allowing high signal-to-noise measurements of the largest scales in a survey. The idea behind cosmic variance cancellation is that observations of different tracers of large-scale structure sample the same realization of the matter density field. A mode-by-mode comparison of the different tracers can therefore avoid sample variance.

As evidenced by Eq.~\eqref{eq:squeezed-kSZ}, the kSZ bispectrum measures the product of $P_{\mathrm{X}e}$ and $P_{\mathrm{X}v}$. A factor can therefore be exchanged between power spectra while keeping the total signal constant. This potential source of confusion is called the optical depth degeneracy. When reconstructing the velocity field, the uncertainty on $P_{\mathrm{X}e}$ can be accounted for through the parameter $b_v$, the velocity reconstruction bias, which we include in our modelling.

We consider the data vector $\boldsymbol{\delta}$, built as the concatenation of the reconstructed velocity field and the perturbations in the X field. That is, $\boldsymbol{\delta} = \left\lbrace v,\delta_{\rm{X}} \right\rbrace(k,\mu,z)$, where $\mu\equiv k_{r}/k$ can take values from $-1$ to 1. The covariance therefore depends on the auto-power spectrum of the overdensities of the tracer $X$, of the velocity field $v$, and their cross-correlation: $P_{\rm XX}$, $P_{vv}$, and $P_{{\rm X}v}$, respectively. The signal and noise covariances of the X and $v$ perturbations can be expressed as
\begin{equation}
\begin{split}
\mathbf{S}(k,\mu,z) &= \begin{pmatrix} P_{vv} & P_{{\rm X}v} \\ P_{{\rm X}v} & P_{{\rm XX}}\end{pmatrix}, \\ \mathbf{N}(k,\mu,z) &= \begin{pmatrix} N_{vv} & 0 \\ 0 & N_{{\rm XX}}\end{pmatrix},
\end{split}
\end{equation}
respectively. The full covariance matrix is then given by the sum of the signal and noise covariances
\begin{equation}
    \mathbf{C}(k,\mu,z) = \mathbf{S}(k,\mu,z) + \mathbf{N}(k,\mu,z).
\end{equation}

Finally, for a single redshift bin and a set of cosmological parameters $\pmb{\theta}$, the Fisher matrix element 
can be written as
\begin{equation}
F_{\alpha \beta} = V\int\int\frac{k^2 dk\, d\mu}{8\pi^2}{\rm Tr} \left[\frac{\partial \mathbf{C}}{\partial \theta_{\alpha}} \mathbf{C}^{-1} \frac{\partial \mathbf{C}}{\partial \theta_{\beta}} \mathbf{C}^{-1} \right].
\label{eq:fisher}
\end{equation}

\section{Clustering Model}\label{sec:LIM}
The tracer of large-scale structure for kSZ tomography we consider in this work is the intensity fluctuation of a given spectral line. Since dark matter halos host star-forming galaxies that source the line emitters, we assume a relation between line intensity and halo mass. We can therefore compute the LIM power spectrum using the halo model \cite{COORAY20021}. In this framework, the power spectrum is modelled as a sum of the correlation within a single halo and between different halos, which correspond to the one-halo ($P_{1h}$) and two-halo ($P_{2h}$) terms, respectively. For two different tracers $X$ and $Y$, we have that

\begin{equation}
\begin{split}
P_{1h}^{\mathrm{XY}}(k,z) &= \int d M \frac{dn}{dM} F_{\mathrm{X}}(k, M, z) F_{\mathrm{Y}}(k, M, z) \\
P_{2h}^{\mathrm{XY}}(k,z) &= \left[\int dM \frac{dn}{dM} b_{\rm X}(M,z) F_{\mathrm{X}}(k, M, z)\right]\times \\ &\times \left[\int dM \frac{dn}{dM} b_{\rm Y}(M,z) F_{\mathrm{Y}}(k, M, z)\right] P_{\mathrm{lin}}(k,z),
\end{split}
\end{equation}
where $M$ is the halo mass, $dn/dM$ is the halo mass function, $b_{\rm X}$ is the weighted linear halo bias for the tracer X (weighted by the luminosity of the line in the case of intensity maps), $F_{\mathrm{X}}$ is a profile function that depends on the tracer X, and $P_{\rm lin}$ is the linear matter power spectrum. We use the halo bias fitting function and the halo mass function from Ref.~\cite{Tinker_hmf2010}. In this work, we apply the halo model formalism to compute the auto and cross-power spectra for CII intensity and electron clustering.

The discreteness of the line emitters leads to an additional contribution to the LIM auto-power spectrum: the shot-noise. This scale-invariant term is given by
\begin{equation}
    P^{\mathrm{X}}_{\mathrm{shot}} (z) = \left[\frac{c }{4\pi \nu H(z)}\right]^2 \int dM\, L_{\mathrm{X}}^2(M,z) \frac{dn}{dM},
\end{equation}
where $c$ is the speed of light, $\nu$ is the emission frequency of the spectral line, and $L_{\rm X}$ is the luminosity of the line $X$ associated to a halo of mass $M$. The full LIM signal is therefore given by the sum of the clustering and shot-noise terms, i.e.
\begin{equation}
    P^{\mathrm{XY}}(k,z) = P_{1h}^{\mathrm{XY}}(k,z) + P_{2h}^{\mathrm{XY}}(k,z) + \delta^{\mathrm{XY}}_K P^{\mathrm{X}}_{\mathrm{shot}}(z).
    \label{eq:halomodel}
\end{equation}
where $\delta_K$ is the Kronecker delta.

The expression above describes the intrinsic LIM power spectrum signal. However, the \textit{observed} signal is limited by instrumental white noise as well as  the resolution and finite volume of the survey. The latter two effects limit the access to small and large scale modes, respectively. We implement this loss of observed modes by multiplying each power of LIM fluctuations in Eq.~\eqref{eq:halomodel} by the square root of the total window function $W(k,\mu,z)$ (see Ref.~\cite{Bernal:2019jdo} for further details).
Residual foreground contamination and line interlopers are additional sources of noise, but we neglect their contribution in this work.

Assuming a uniform observation of the survey volume, where each voxel volume element $V_{\mathrm{vox}}$, covering an area $\Omega_{\rm {pix}}$, is observed over a time $t_{\mathrm{pix}}$, then the LIM instrumental noise power spectrum can be modelled as
\begin{equation}
    N_{\mathrm{X}} = \sigma^2_{\rm{N}} V_{\rm{vox}} = \frac{\sigma^2_{\rm{pix}}}{t_{\rm{pix}}} V_{\mathrm{vox}},
    \label{eq:Pn}
\end{equation}
where $\sigma_{\rm{pix}}$ is 
typically given in terms of a noise equivalent intensity (NEI), in units of Jy s$^{1/2}$/sr, and $\sigma_{\rm{N}}$ is the final survey sensitivity per voxel.
The NEI can be converted to a noise equivalent flux density (NEFD) using the solid angle of the pixel (taken here to be given by the solid angle of a Gaussian beam $\Omega_{\rm{pix}} = \Omega_{\rm {beam}}=[\sigma_{\rm beam}^X]^2=[\theta_{\rm FWHM}^X]^2/8\ln 2$); that is, $\rm{NEI} = \rm{NEFD}/\Omega_{\rm{pix}}$. The observing time per pixel is computed as $t_{\mathrm{pix}}=t_{\mathrm{obs}}/N_{\mathrm{pix}}$, where $t_{\mathrm{obs}}$ is the total observing time, and $N_{\mathrm{pix}} = \Omega_{\mathrm{field}}/\Omega_{\mathrm{beam}}$ is the number of pixels.

\section{Potential of kSZ with LIM}\label{sec:SNR}
\begin{table*}
\centering
 \begin{tabular}{c c c c c c c c}
 \hline
 Experiment & Frequencies & FWHM  & Resolving power & Survey area & Observing time & Variance per voxel & Noise power spectrum\\
 & (GHz) & (arcsec) &  & (deg$^2$) & (h) & (Jy sr$^{-1}$) & (Mpc$^3$ Jy$^2$ sr$^{-2}$) \\
 \hline\hline
 CCAT-prime & $388-428$ & 30 & 100 & 8 & 4000 &  $5.7\times 10^4$ & $2.3\times 10^{10}$ \\
 AtLAST-like & $320-950$ & 8 & 1000 & 7500 & 10000 & $5.6\times 10^{4}$ & $1.5\times 10^8$\\
 \hline
 \end{tabular}
  \caption{Experimental parameters for the LIM instruments considered in this work. We report in this table the noise white power spectrum at $z=3.5$, with a bin width of $\Delta z=1$ for the AtLAST-like survey.}
    \label{tab:tabCII_exp}
\end{table*}
In this section we apply the formalism outlined above to CII intensity mapping. We begin by describing the model we adopt for the clustering of the CII line and provide details for the CII and CMB experiments considered. We then predict the detectability of the kSZ bispectrum signal and of the reconstructed velocity field.

Current observations indicate that CII emissions are primarily originated in photodissociation regions \cite{2012A&A...548A..91L, 2014ApJ...781L..15R}. These are regions of the interstellar medium that are predominantly neutral, where the far ultra-violet radiation plays a key role in determining the gas properties. A correlation between CII luminosity and star-formation rate can be established from the relation between CII and far ultra-violet radiation, with far infrared luminosity as an intermediate step. We adopt the model presented in Ref.~\cite{Silva:2014ira}, which assumes CII emissions are sourced within dark matter halos and parametrizes the relation between CII and star-formation rate, converting it into a relation between CII and halo mass.

With the CII luminosity-halo mass relation at hand, we can proceed to calculate the relevant power spectra: the CII auto-correlation and the cross-correlation between the CII intensity and the electron distribution. We assume the CII intensity follows a Navarro-Frenk-White (NFW) density profile \cite{Navarro:1996gj} using the halo concentration-mass relation from Ref.~\cite{Diemer_cNFW}, and that the electron gas follow the AGN feedback profile given in Ref.~\cite{Battaglia:2016xbi}. For each tracer, we have
\begin{equation}
    \begin{split}
        F_{\mathrm{CII}}(M,z) &= \frac{c}{4\pi \nu H(z)} L_{\mathrm{CII}}(M,z) u_{\mathrm{NFW}}(k, M, z), \\
        F_e(M,z) &= \frac{M}{\bar{\rho}_m} u_{\mathrm{gas}}(k,M,z),
    \end{split}
\end{equation}
where $\bar{\rho}_m$ is the present-day mean matter density and $u$ is the Fourier transform of the density profile.

\subsection{Detectability}
We assess the detectability of the cosmological kSZ signal in the near future with current and upcoming CMB and LIM surveys. To do so, we compute the signal-to-noise ratio $S/N$ of the measurement of the kSZ bispectrum using Eq.~\eqref{eq:SNR}. We consider the following CII experiments:
\begin{itemize}
    \item The Epoch of Reionization Spectrometer (EoR-Spec) on CCAT-prime is designed to probe CII emissions at high-redshift \cite{2020JLTP..199.1089C}. The survey spans the redshift range $z = 3.5-8.1$, and covers $\Omega_{\mathrm{field}} = 8$ deg$^2$ of the sky over an observing time of $t_{\mathrm{obs}} = 4000$ h . Here we consider observed frequencies between $\nu_{\rm{obs}} = 388-428$ GHz, with a resolving power of $R = \delta_{\nu}/\nu_{\rm{obs}}=100$, where $\delta_{\nu}$ is the spectral resolution.
    \item The Atacama Large Aperture Submillimeter Telescope (AtLAST) is a proposed next-generation 50~m single dish telescope \cite{2019BAAS...51g..58K}. Its goal is to achieve a combination of angular and spectral resolution, as well as sensitivity and mapping speed, necessary for probing large-scale structure. The experiment aims to cover the entire SDSS field of view of $\Omega_{\mathrm{field}} = 7500$~deg$^2$. We consider the frequency range $320-950$ GHz, corresponding to redshifts $z=1-5$ for the CII line.
\end{itemize}

The instrumental parameters used for the experiments above are summarized in Table~\ref{tab:tabCII_exp}. We consider a single redshift bin for CCAT-prime centered at $z=3.7$, with width $\Delta z = 0.5$. For the AtLAST-like survey, we divide the total frequency range in 4 bins of width $\Delta z = 1$. In addition to the instrument design, we report the total variance per voxel and the white noise power spectrum, defined in Eq.~(\ref{eq:Pn}). Since the AtLAST-like configuration encompasses multiple redshift bins, we report only the value of the noise power spectrum for the bin centered at $z=3.5$. The beam FWHM is set to a conservative value of $\theta_{\rm{FWHM}} = 8$~arcsec.

The two LIM experiments described above are then cross-correlated with SO and CMB-S4. We consider  $\theta^{\rm{CMB}}_{\mathrm{FWHM}} = 1.4$ arcmin and $s = 6\ \mu {\rm K}$-arcmin for the SO configuration. Although a precise instrument specification is yet to be defined for CMB-S4, we assume $\theta^{\rm{CMB}}_{\mathrm{FWHM}} = 1$ arcmin and $s = 1\ \mu {\rm K}$-arcmin.

\begin{figure}[t]
    \centering
    \includegraphics[width=0.48\textwidth]{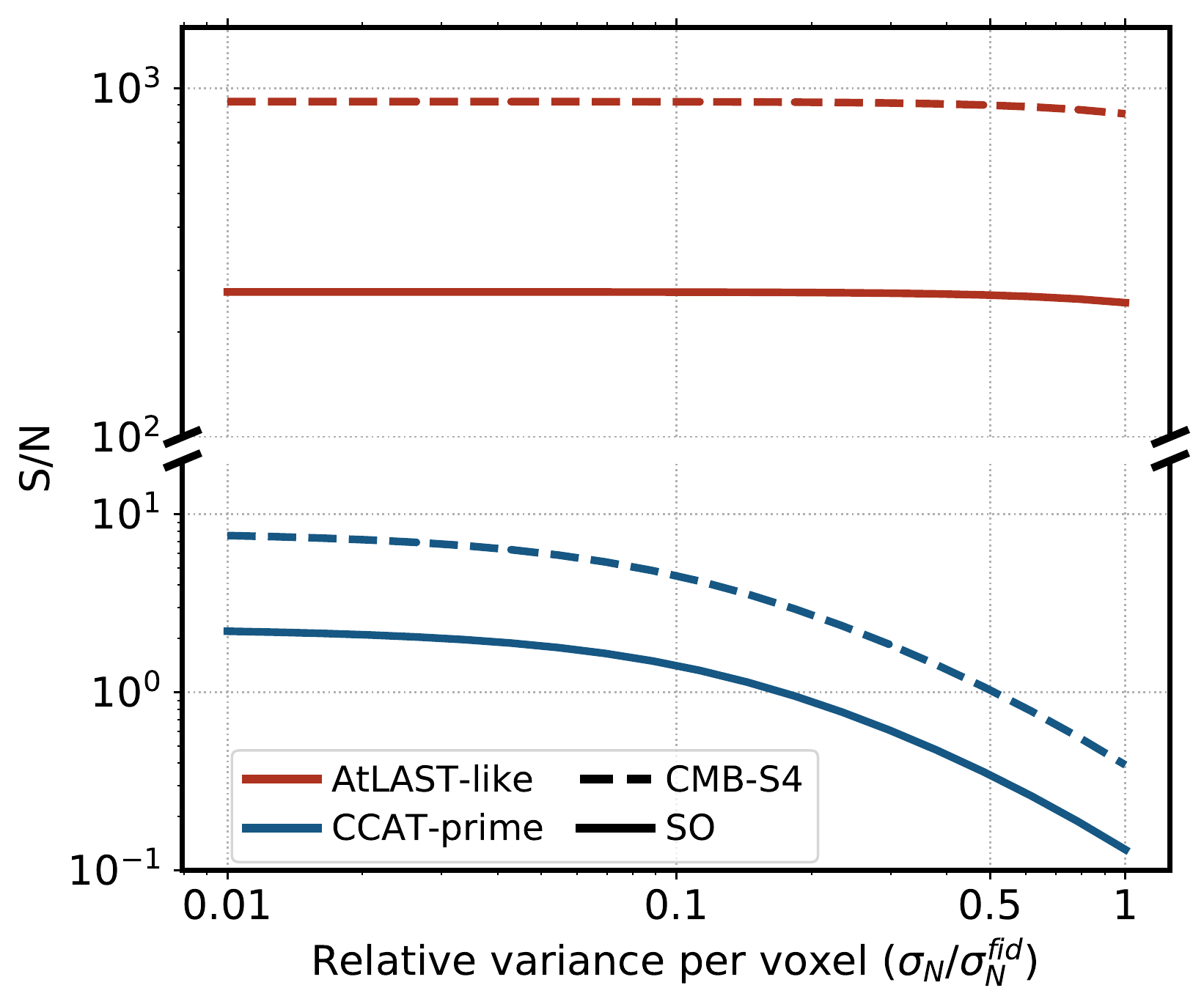}
    \caption{Signal-to-noise ratio of the total kSZ tomography signal as a function of the variance per voxel ($\sigma_{\rm{N}}$) in the LIM survey, relative to the fiducial value ($\sigma^{\rm{fid}}_{\rm{N}}$) given in Table~\ref{tab:tabCII_exp}. We consider the cross-correlations between the CII intensity mapping surveys CCAT-prime and AtLAST-like, and the CMB experiments SO and CMB-S4.}
    \label{fig:SNR}
\end{figure}

Figure~\ref{fig:SNR} shows the total signal-to-noise ratio of the kSZ bispectrum measurements using CII intensity mapping, summed over all wavenumbers and, for the AtLAST-like survey, over all redshift bins. The resulting sensitivity is shown as a function of relative changes in $\sigma_{\rm{N}}$ for each of the 4 experiment combinations, where 1 corresponds to the fiducial value. We show that a $1\sigma$ measurement of kSZ tomography at $z\sim 3.7$ with the spectroscopic-imaging module on CCAT-prime in cross-correlation with SO would require the LIM noise power spectrum to be reduced by a factor of 5. With CMB-S4, a reduction of the instrumental noise by a factor of 2 suffices for a 1$\sigma$ detection.

In the forecasted $S/N$ shown in Fig.~\ref{fig:SNR}, we assumed the first-light experimental design for CCAT-prime. We note, however, that the instrument has substantial upgrade potential. The current Phase I design has the possibility of accommodating seven instrument modules \cite{Vavagiakis:2018gen}, with further extensions (e.g. Phase II) housing up to 19 optics tubes. We also consider a ``Phase II" instrument design, with a field of view of $\Omega_{\rm{field}} = 16$~deg$^2$, a longer observing time of $t_{\rm{obs}} = 10000$~h, 7 spectrometers, a higher resolving power of $R=300$, and a total variance per voxel of $\sigma_{\rm{N}} = 5\times 10^3$ Jy sr$^{-1}$. Under this configuration, we find that a $3\sigma$ detection of the kSZ signal can be achieved at $z\sim 3.7$ in cross-correlation with SO.

While instruments that are currently under construction may offer a modest but unprecedented kSZ detection at high redshifts, we show that a high signal-to-noise detection [$\mathcal{O}(10^2-10^3)$] of the kSZ effect can be achieved with a wide-field LIM survey such as AtLAST. These high-precision measurements will open the possibility for exploiting the kSZ effect measured with LIM for cosmological analyses, thanks to a powerful reconstruction of the velocity field at large scales and high redshift.

\begin{figure*}[t]
    \centering
    \includegraphics[width=0.8\textwidth]{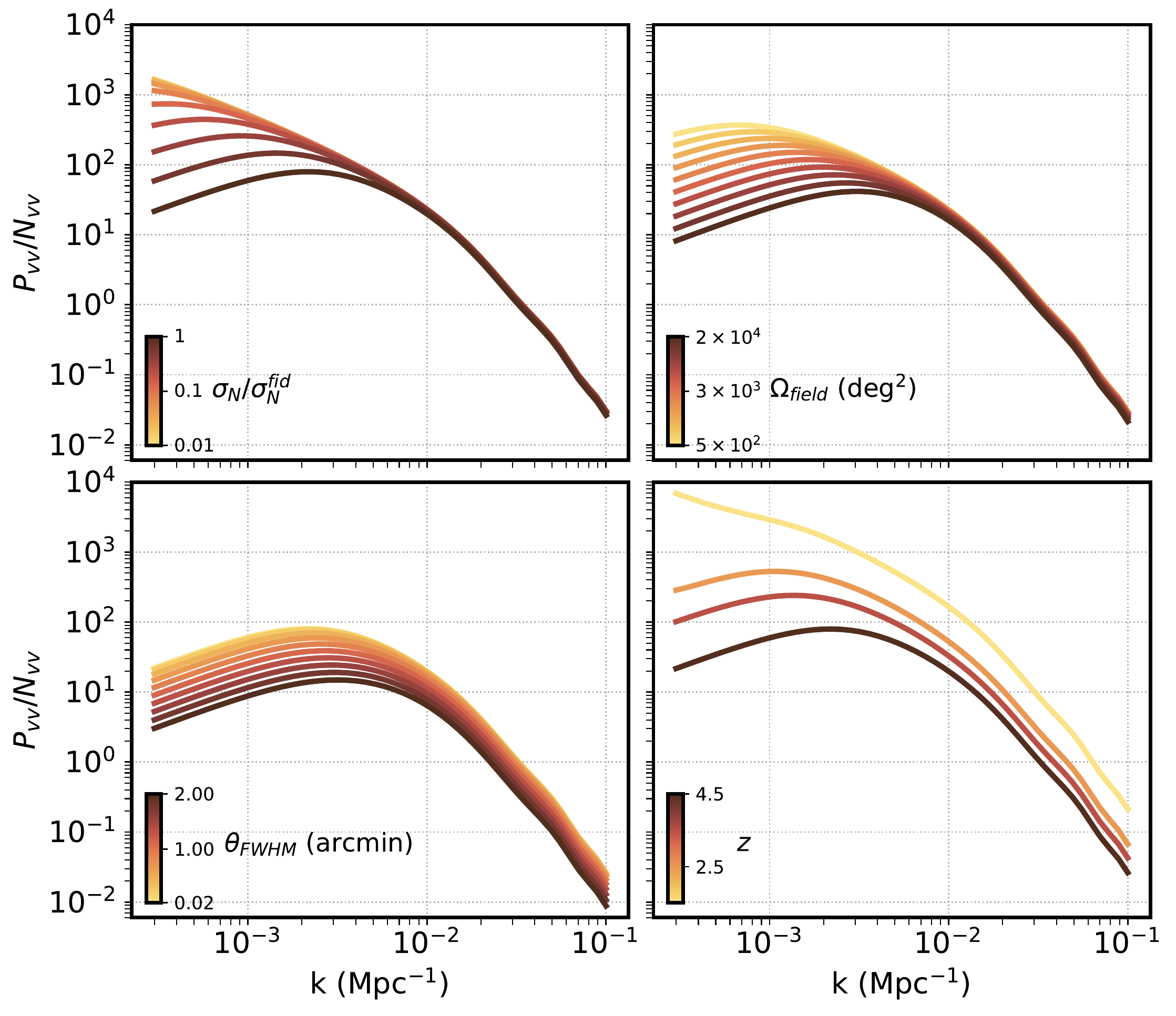}
    \caption{Signal-to-noise ratio of the reconstructed velocity power spectrum, where the reconstruction noise is computed using Eqs.~\ref{eq:Nvr} and \ref{eq:Nvv}. Each panel shows  $P_{vv}/N_{vv}$ as a function of scale and how it depends on a given parameter of the LIM instrument. The upper left corresponds to the noise sensitivity per voxel, the upper right to the survey area, the lower left to the FWHM of the beam profile full-width half maximum of the LIM experiment, and the lower right to the redshift.}
    \label{fig:vel_rec}
\end{figure*}

\subsection{Velocity Reconstruction}\label{sec:kSZxLIM}
Here we envision the next generation of LIM surveys and forecast their potential to infer cosmological parameters across cosmic times using kSZ tomography. In this section we consider the cross-correlation between CMB-S4 with a wide-field single-dish telescope based on AtLAST.

The benefit of combining the reconstructed velocity field with the LIM power spectrum stems from sample-variance cancellation. This reduction of sample variance boosts the significance of observations at large scales (i.e., those that are sample-variance limited). In order to achieve a precise velocity reconstruction at large scales, high sensitivity to both large and small scales is required, since the kSZ effect relies on physics on both regimes. For LIM experiments, this requirement translates to surveying large volumes and having sufficient angular and spectral resolution.

We show the signal-to-noise ratio ($P_{vv}/N_{vv}$) of the power spectrum of the reconstructed velocity field as function of scale Fig.~\ref{fig:vel_rec}. In each panel, we show how $P_{vv}/N_{vv}$ depends on certain experimental parameters of interest. Logarithmic bins of width $\Delta(\log_{10}k) = 0.05$ have been assumed. In what follows, we discuss some of the main features observed in Fig.~\ref{fig:vel_rec}.

High $\sigma_{\rm{N}}$ leads to a noise power spectrum that dominates over the LIM power spectrum at small scales, which translates to a reduction of the $P_{vv}/N_{vv}$ at large scales. In a similar way, as $\Omega_{\rm field}$ increases, so does the noise power spectrum (for a fixed $\sigma_{\rm N}$). This effect dominates over the reduction of cosmic variance; therefore, increasing $\Omega_{\rm field}$ presents a similar behavior to increasing $\sigma_{\rm N}$. Reducing $\theta_{\rm FWHM}$ increases the number of voxels and reduces their volume, which results in a smaller noise power spectrum for a fixed $\sigma_{\rm N}$. Moreover, it also improves the resolution of the survey, granting access to smaller scales, so that the $P_{vv}/N_{vv}$ also improves for large $k$. Finally, the $P_{vv}/N_{vv}$ decays with redshift because the overall amplitude of the LIM power spectrum decreases due to amplitude of the matter perturbations (even if the CII luminosity peaks at $z\sim 2$). $P_{vv}/N_{vv}$ at large scales degrades faster with redshift due to the loss of modes along the line of sight, since the size of the volume probed falls with redshift for bins with the same width in redshift.

This study shows how the modes that can be reconstructed depend on the instrumental parameters of the LIM survey and the redshift at which it is being observed. This study highlights the impact of both large and small scale sensitivity in the LIM survey. Furthermore, it can clarify what survey design is required to achieve different scientific goals using kSZ tomography.

\section{Case study: PNG and CIPs}\label{sec:PNG&CIP}
The possibility of testing models of the primordial Universe is one of the most compelling motivations to probe clustering on the largest scales. Here we consider inflationary theories that feature local-type PNGs, parametrized by $f_{\rm NL}$, and CIPs, parametrized by $A_{\rm CIP}$, that are fluctuations of baryons and cold dark matter that leave the total matter distribution unchanged and are correlated with adiabatic perturbations. Both signatures of inflation leave an imprint on the halo bias that scales as $1/k^2$.

\subsection{Model}\label{sec:PNG&CIP_model}
In order to assess the gain in information on cosmological parameters offered by adding kSZ velocity reconstruction to the analysis of a future LIM survey, we compute the Fisher information matrix given in Eq.~\eqref{eq:fisher}. We include the contribution from both PNG of the local type and correlated CIPs in our power spectrum model, as well as the effect of redshift-space distortions.

The relevant power spectra are then given by
\begin{equation}
\begin{split}
P_{\mathrm{XX}}&(k,\mu,z) = \left[ b_{\rm X}(z) + f\mu^2 + f_{\rm NL} \Delta b^{f_{\mathrm{NL}}}_{\rm X}(k,z) + \right. \\ &+ \left. A_{\rm CIP} \Delta b^{A_{\mathrm{CIP}}}_{\rm X} (k,z)\right]^2 T^2_{\mathrm{X}}(z) W(k,
\mu, z) P_{\rm{lin}}(k,z)\\
P_{v\mathrm{X}}&(k,\mu,z) = \left(\frac{b_v faH}{k}\right) \left[ b_{\rm X}(z) + f\mu^2 + \right. \\ &+ \left. f_{\rm NL} \Delta b^{f_{\rm NL}}_{\rm X}(k,z) + A_{\rm CIP} \Delta b^{A_{\rm CIP}}_{\rm X} (k,z)\right]\times \\ &\times I_{\mathrm{X}}(z) \sqrt{W(k,
\mu, z)} P_{\rm{lin}}(k,z)\\
P_{vv}&(k,z) = \left(\frac{b_v faH}{k}\right)^2 P_{\rm{lin}}(k,z)
\end{split}
\end{equation}
where the Gaussian bias $b_{\rm X}$ is modulated by the scale-dependent biases induced by PNGs ($\Delta b^{f_{\mathrm{NL}}}_X$) and CIPs ($\Delta b^{A_{\mathrm{CIP}}}_X$), and $I_X$ is the mean specific intensity of the spectral line of interest. We note that only the linear matter power spectrum is considered, since the reconstructed modes are restricted to large scales. The scale-dependent bias contributions have been defined as \cite{Hotinli:2019wdp, Barreira:2019qdl, Barreira:2020lva}
\begin{equation}
\label{eq:PS_fisher}
    \begin{split}
        \Delta b^{f_{\mathrm{NL}}}_{\rm X}(k,z) &= \left(b_{\rm X} -1 \right) \delta_{\rm{cr}} \frac{3\Omega_m H^2_0}{k^2 T_{m}(k,z)}, \\
        \Delta b^{A_{\mathrm{CIP}}}_{\rm X}(k,z) &= b_{bc}(z)                                                                                                        f_{bc} \frac{5\Omega_m H^2_0}{2 k^2 T_{m}(k, z)},
    \end{split}
\end{equation}
where $\delta_{\rm{cr}}=1.68$ is the critical density for spherical collapse, $\Omega_m$ is the matter density parameter at $z=0$, $T_m(k,z)$ is the matter transfer function, and we define $f_{bc}\equiv 1+\Omega_b/\Omega_c$, where $\Omega_b$ and $\Omega_c$ are the density parameters of baryon and cold dark matter today, respectively. The quantity $b_{bc}(z)$ is the bias coefficient corresponding to the CIP contribution to the halo density perturbation, which was fit to be \cite{Hotinli:2019wdp}
\begin{equation}
b_{bc}(z) = -(0.16 + 0.2z + 0.083z^2).
\end{equation}

\begin{figure}[t]
    \centering
    \includegraphics[width=0.45\textwidth]{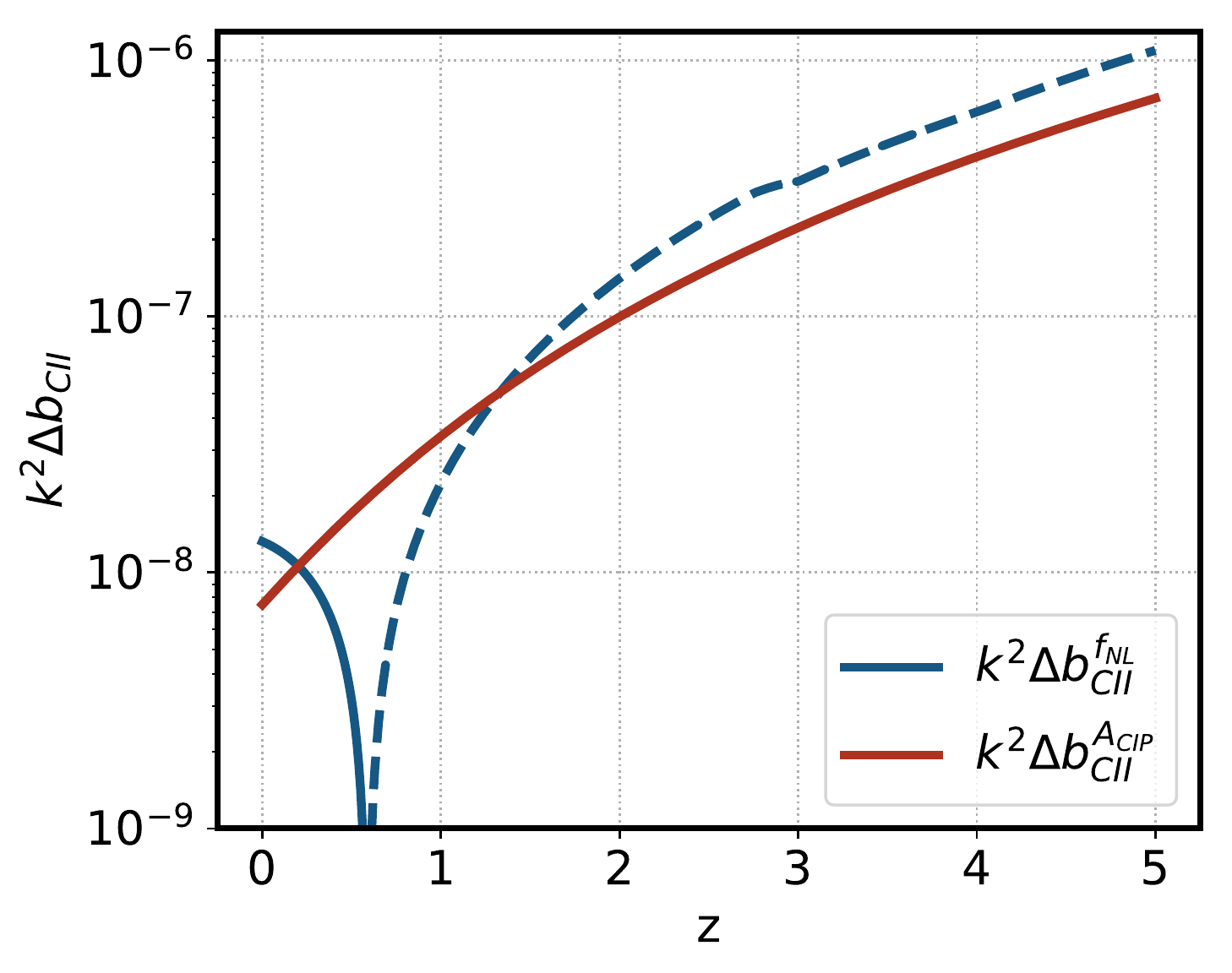}
    \caption{Change in the halo bias induced by local-type PNG and CIPs as a function of redshift. The solid lines correspond to negative values and the dashed to positive.}
    \label{fig:Db}
\end{figure}

Figure~\ref{fig:Db} shows the redshift dependence of the bias modulation $k^2 \Delta b_{\rm CII}$ on large scales for both the PNG and CIP contributions. The redshift evolution of the two contributions only differ significantly for $z \lesssim 1.5$, where $\Delta b^{f_{\rm{NL}}}_{\rm CII}$ changes sign at $z\sim 0.6$. We note that the aforementioned features in $\Delta b_{\rm CII}$ depend on the spectral line under study, since the bias term $b_{\rm X}$ in Eq.~\eqref{eq:PS_fisher} is a luminosity-weighted bias. This dependence and its impact on the redshift evolution of $\Delta b^{f_{\rm NL}}$ can be used to further break the degeneracy between $f_{\rm NL}$ and $A_{\rm CIP}$. Furthermore, Fig.~\ref{fig:Db} shows that the effect of PNGs and CIPs grow significantly at higher redshifts.  We therefore expect measurements at high redshifts to offer tighter constraints of $f_{\rm{NL}}$ and $A_{\rm{CIP}}$, albeit highly degenerate. The degeneracy can then be broken by including a low-redshift measurement.

\begin{figure*}[t]
    \centering
    \includegraphics[width=0.9\textwidth]{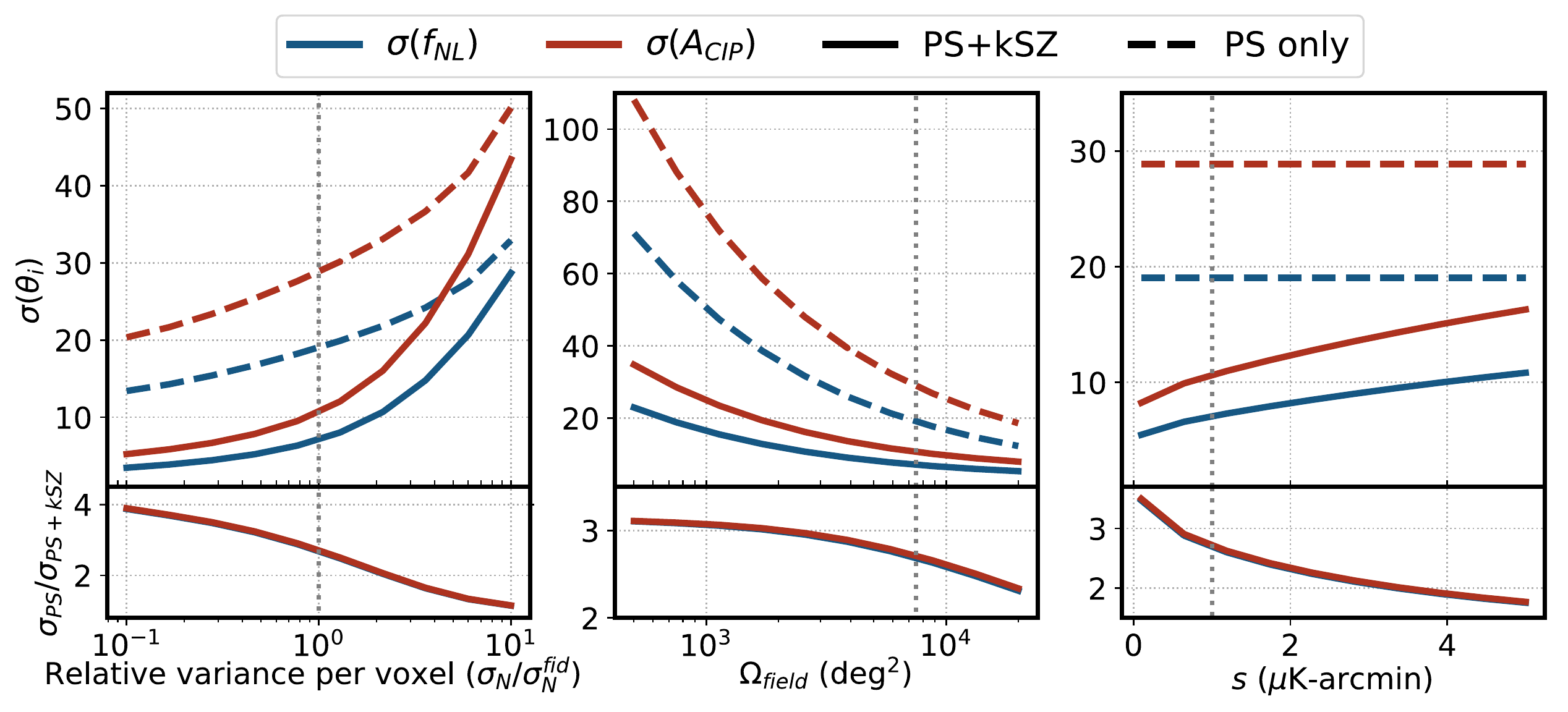}
    \caption{Marginalized 68\% confidence-level forecast on $f_{\rm{NL}}$ and $A_{\rm{CIP}}$ as a function of experimental parameters. Forecasts assuming only a measurement of the LIM power spectrum are shown in dashed lines, and forecasts combining both the LIM power spectrum and kSZ tomography are shown in solid lines. The vertical dotted lines correspond to the fiducial intrument parameters. The left panel shows how the constraints depend on the sensitivity per pixel of the LIM survey, the middle panel shows the dependence on the overlapping field of view, and the right panel shows the dependence on the CMB sensitivity. Notice the change in scales in the y-axis for each panel. The forecasts include the redshift range $z=1-5$, divided in bins of width $\Delta z=1$.}
    \label{fig:sigma_instr}
\end{figure*}

\subsection{Forecasts}
The CMB and LIM instruments we consider in our forecasts are CMB-S4 and the AtLAST-like survey, with the fiducial parameters given in Sec.~\ref{sec:SNR}. We include the potential presence of both local-type PNG and CIPs, unless otherwise stated. The complete set of parameters considered here are:  $h$, $\Omega_m$, $\Omega_b$, $n_s$, $A_s$, $f_{\rm{NL}}$, $A_{\rm{CIP}}$, and $b_v$. The results presented are marginalized over all standard $\Lambda$CDM parameters and velocity reconstruction bias.

We show in Fig.~\ref{fig:sigma_instr} how the uncertainty on $f_{\mathrm{NL}}$ and $A_{\mathrm{CIP}}$ depend on LIM and CMB instrumental parameters. We report the forecasted marginalized 68\% confidence level uncertainties for both a study that only considers a measurement of the CII intensity fluctuations and one that additionally includes the reconstructed velocity field. The results shown include the redshift range $z=1-5$, divided in bins of width $\Delta z = 1$.

We show in Table~\ref{tab:fNL_ACIP_results} the constraints on $f_{\rm{NL}}$ and $A_{\rm{CIP}}$ for different models. With both PNG and CIPs, assuming the fiducial set of instrumental parameters, we find that the CII experiment alone can achieve an uncertainty of $\sigma(f_{\mathrm{NL}}) = 19$ and $\sigma(A_{\mathrm{CIP}})=29$. By including the reconstructed velocity field from kSZ tomography, we find $\sigma(f_{\mathrm{NL}}) = 7$ and $\sigma(A_{\mathrm{CIP}})=10$, both roughly corresponding to an improvement by a factor of 3. The left panel of Fig.~\ref{fig:sigma_instr} shows that cosmic-variance cancellation becomes more pronounced as the LIM instrumental noise is reduced due to a more precise velocity reconstruction.

The first column in Table~\ref{tab:fNL_ACIP_results} corresponds to the more commonly adopted model where only local-type PNG is considered and $A_{\mathrm{CIP}}$ is fixed to zero. We show that the standard observational goal of $\sigma(f_{\rm{NL}}) \lesssim 1$ may be within reach for the next generation of wide-field LIM experiments and that the inclusion of kSZ tomography can be a significant step to achieve this limit. The second column shows the analogous result for $A_{\mathrm{CIP}}$, when $f_{\mathrm{NL}}$ is fixed to zero.

\begin{table}[t]
\centering
\resizebox{0.7\columnwidth}{!}{%
\begin{tabular}{|c|c|c|c|c|}
\hline
Model & PNG & CIP & \multicolumn{2}{c|}{PNG+CIP} \\ \hline
Parameter & $f_{\rm NL}$ & $A_{\rm CIP}$ & $f_{\rm NL}$ & $A_{\rm CIP}$ \\ \hline \hline
PS & 1.5 & 2.3 & 19 & 29 \\ \hline
PS+kSZ & 0.7 & 1.1 & 7 & 10 \\ \hline
\end{tabular}%
}
\caption{Forecasted marginalized 68\% confidence-level constraints on $f_{\rm{NL}}$ and $A_{\rm{CIP}}$ for the different models and combination of observables considered. The first row corresponds to forecasts assuming only a measurement of the LIM power spectrum. In the second row, we show forecasts for a combination of both the LIM power spectrum and kSZ tomography.}
\label{tab:fNL_ACIP_results}
\end{table}

\begin{figure*}[t]
    \centering
    \includegraphics[width=0.8\textwidth]{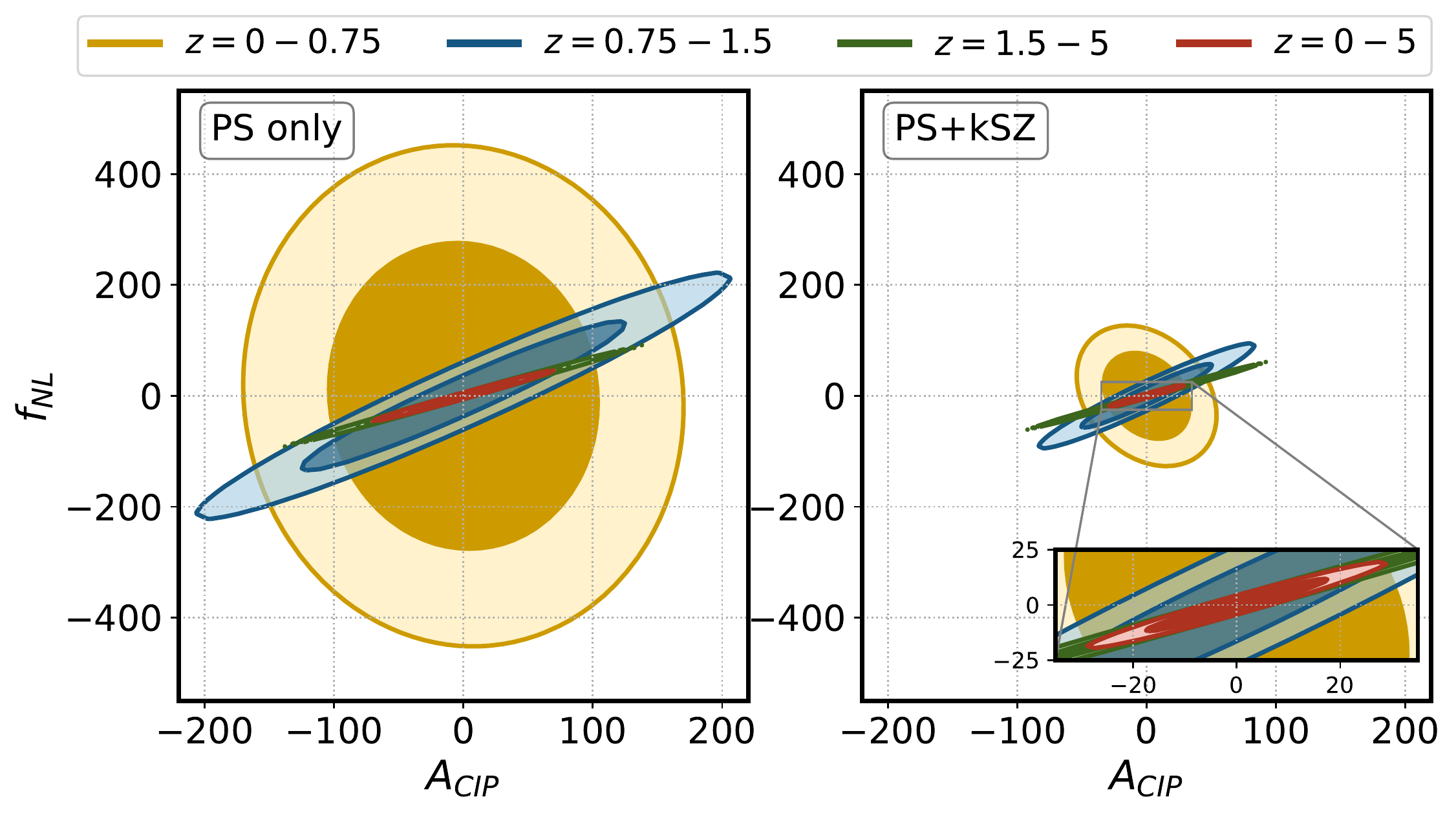}
    \caption{Forecasted uncertainties on $f_{\mathrm{NL}}$ and $A_{\mathrm{CIP}}$ at 68\% and 95\% confidence levels, marginalized over cosmological and bias parameters. The panel on the left shows the forecast when only the LIM survey is considered, i.e., the constraint obtained only from the CII power spectrum. The panel on the right shows the result when the reconstructed velocity field is added to the analysis. Each color corresponds to the redshift range included in the forecast.}
    \label{fig:fisher_zs}
\end{figure*}

In order to explore the degeneracy between $f_{\rm{NL}}$ and $A_{\rm{CIP}}$, we show in Fig.~\ref{fig:fisher_zs} the constraints as a function of redshift. We expand the redshift range of the LIM survey to $z = 0-5$ in order to capture the sign change in the bias modulation described in Section~\ref{sec:PNG&CIP_model}. Since the parameters are perfectly degenerate for a single redshift bin, we consider thinner bins of $\Delta z = 0.25$ and add the corresponding Fisher matrices in the intervals displayed in Fig.~\ref{fig:fisher_zs}.

Figure~\ref{fig:fisher_zs} shows that the degeneracy between $f_{\rm{NL}}$ and $A_{\rm{CIP}}$ differs for low ($z\lesssim 0.75$) and high ($z\gtrsim 1.5$) redshifts. While forecasts at redshifts above $\sim 1.5$ show tighter constraints, including lower redshifts is essential to break the degeneracy between the two parameters, evident from Fig.~\ref{fig:Db}. Note that our choice of thinner redshift bins in Fig.~\ref{fig:fisher_zs} leads to the loss of large-scale radial modes, resulting in weaker constraints than those reported in Table~\ref{tab:fNL_ACIP_results} and Fig.~\ref{fig:sigma_instr}.

\section{Conclusion}\label{sec:conclusion}

Line-intensity mapping has emerged as a new observational technique capable of
offering a window into large cosmological volumes beyond the reach of galaxy surveys. The prospect of cross-correlating intensity maps with CMB observations has recently garnered attention, due to its potential to probe large-scale structure across a wide range of redshifts. In this work, we consider the measurement of kSZ tomography in the post-reionization Universe using LIM.

We predict the expected sensitivity to the kSZ signal for different combinations of planned LIM and CMB experiments. Here we focus on CII intensity mapping, but emphasize that the technique can be applied to any atomic or molecular line. We consider two generations of experiments: those currently under construction expected to operate in the next few years (CCAT-prime and SO) and next-generation experiments planned for the upcoming decade (AtLAST and CMB-S4).

We show that while the first-light instrument design for CCAT-prime is unable to measure kSZ tomography at $z \sim 3.7$ in cross-correlation with SO, a low significance detection may be within reach for an upgraded ``Phase II" design. Furthermore, the detection significance will be substantially increased through the next generation of experiments, with a signal-to-noise ratio of roughly $\mathcal{O}(10^2-10^3)$ for the cross-correlation between CMB-S4 and an AtLAST-like survey.

High-significance measurements of kSZ tomography in cross-correlation with the next generation of wide-field LIM surveys will enable the large-scale velocity field to be recovered with high fidelity. Three-dimensional maps of the velocity field may therefore be extended to redshifts beyond the reach of planned galaxy surveys, offering information about gravity and matter clustering at high redshifts. We provide a study of how the signal-to-noise ratio of the power spectrum of reconstructed velocities depends on the LIM survey parameters, which provides insight to optimize the design of LIM experiments.
As a case example, we estimate the performance of kSZ tomography with LIM to constrain models of inflation through its impact on clustering on large scales. Both PNG and CIPs induce a contribution to the halo bias with the same scale dependence, although with differing redshift evolution.

We show that while next-generation LIM instruments alone may offer competitive constraints on inflation, including the velocity field reconstructed from kSZ tomography can significantly boost these constraints. With LIM and CMB experiments based on AtLAST and CMB-S4, we show that the inclusion of the reconstructed velocity field can improve constraints on $f_{\rm NL}$ and $A_{\rm CIP}$ by a factor of $\sim 3$. We further explore how these constraints depend on the chosen instrumental parameters and show that the LIM noise power spectrum is the main limitation to velocity reconstruction.

We then investigate the source of the broken degeneracy between $f_{\rm NL}$ and $A_{\rm CIP}$. We compute the constraints on both parameters as a function of redshift and show that the degeneracy at low redshifts ($z\lesssim$ 0.75) and at high redshifts ($z\gtrsim 1.5$) are different in our case of study. While we show that higher redshifts can offer tighter constraints, including low redshifts is essential to break this degeneracy. The need to combine low and high-redshift observations stresses the importance of measuring large-scale structure over a wide redshift range and LIM is a promising technique to achieve this goal.

Recent proposals for next-generation wide-field LIM experiments suggest a promising future for cosmological applications of intensity maps as tracers of large-scale structure. LIM offers a complementary probe to galaxy surveys at low redshifts and is uniquely suited to extend measurements of matter clustering deeper into the observable Universe. Cross-correlations between LIM and the CMB can therefore be critical to fully retrieve the cosmological information present in the secondary anisotropies and to maximize the impact of both LIM and CMB experiments.

\acknowledgments
GSP was supported by the National Science Foundation Graduate Research Fellowship under Grant No.\ DGE1746891.
JLB was supported by the Allan C. and Dorothy H. Davis Fellowship.  This work was supported at Johns Hopkins by NSF Grant No.\ 1818899 and the Simons Foundation.

\appendix
\section{kSZ Tomography bispectrum}\label{app:kSZ}
The fundamental statistical quantity that carries the kSZ tomography signal is the 3-point function involving two powers of the overdensities $\delta_{\mathrm{X}}$, corresponding to a tracer $X$ of large-scale structure, and one power of the integrated temperature fluctuation induced by the kSZ effect $T$, i.e. $\langle \delta_{\mathrm{X}} \delta_{\mathrm{X}} T\rangle$ \cite{Smith:2018bpn}. The kSZ bispectrum $B$ is then defined as
\begin{equation}
\begin{split}
  \langle \delta_{\mathrm{X}} (\pmb{k})\delta_{{\rm X}} (\pmb{k}') T(\pmb{\ell})\rangle =& iB(k,k',\ell, k_r) (2\pi)^3 \times \\ &\times \delta^{(3)}_D\left(\pmb{k} + \pmb{k}' + \frac{\pmb{\ell}}{\chi_{*}}\right),
 \end{split}
\end{equation}
where the subscript `$r$' denotes the component along the line-of-sight, and $\pmb{k}$, $\pmb{k}'$, and $\pmb{\ell}/\chi_{*}$ are wave vectors. Note that the assumed geometry is of a periodic box, where $\delta_{\mathrm{X}} (\pmb{k})$ is defined over 3 spacial dimensions, at a given redshift $z_*$, which corresponds to the comoving distance $\chi_*$, while $T (\pmb{\ell})$ is defined on 2 spacial dimensions.  Notice that the Dirac delta function fixes $k_r = k'_r$, and rotational invariance in the 2D plane implies that the kSZ bispectrum will additionally depend only on the magnitude of the vector wavenumbers $k, k'$, and $\ell$.

We define the bispectrum amplitude estimator as $\hat{\mathcal{E}} \equiv \hat{B}/B$, where $\hat{B}$ is a bispectrum estimator. Since three-point correlation functions are expected to have low signal-to-noise for each individual mode configuration, we sum over all modes with an optimal weight $W(\pmb{k}, \pmb{k}', \pmb{\ell})$. Its most general form is given by
\begin{equation}
\begin{split}
     \hat{\mathcal{E}} =&\int_{d^3\pmb{k}} \int_{d^3\pmb{k}'}\int_{d^2\pmb{\ell}} W(\pmb{k}, \pmb{k}', \pmb{\ell}) \left[ \delta_{\rm X}(\pmb{k}) \delta_{\rm X}(\pmb{k}') T(\pmb{\ell}) \right] \times \\ &\times (2\pi)^3 \delta^{(3)}_D\left(\pmb{k} + \pmb{k}' + \frac{\pmb{\ell}}{\chi_{*}}\right),
     \label{eq:estimator}
\end{split}
 \end{equation}
where we have defined $\int_{d^3\pmb{k}}\equiv \int \frac{d^3\pmb{k}}{(2\pi)^3}$, and $\int_{d^2\pmb{\ell}}\equiv \int \frac{d^2\pmb{\ell}}{(2\pi)^2}$. We impose the constraint  $\langle \hat{\mathcal{E}}\rangle = 1$ for an unbiased estimator.

The optimal weight for the bispectrum estimator (see, e.g., Ref.~\cite{PhysRevD.86.063511}) is given by

\begin{equation}
    W(\pmb{k}, \pmb{k}', \pmb{\ell}) = \frac{1}{2F_{BB}} \frac{-i B^*(k,k',\ell,k_r)}{P^{\rm tot}_{\rm XX}(k) P^{\rm tot}_{\rm XX}(k') C^{\rm tot}_{\ell}}
\end{equation}
where superscript `$*$' denotes the complex conjugation and $F_{BB}$ is the bispectrum Fisher matrix element.

In the limit where all three modes correspond to large scales, the kSZ bispectrum can be approximated by its tree-level contribution, which is given by
\begin{equation}
\begin{split}
    B(k, k', \ell, k_r) =& \frac{K_* k_r}{\chi^2_*}\left[ P_{{\rm X}e}(k)\frac{P_{{\rm X}v}(k')}{k'} - \right. \\ &- \left. \frac{P_{{\rm X}v}(k)}{k} P_{{\rm X}e}(k') \right],
\label{eq:full3pt}
\end{split}
\end{equation}
where $P_{{\rm X}e}$ is the cross-correlation between $\delta_{{\rm X}}$ and the electron density perturbations, $P_{{\rm X}v}$ is the cross-correlation with the velocity field, and we have defined
\begin{equation}
    K_* \equiv -T_{{\rm CMB}} \sigma_T \bar{n}_{e,0} e^{-\tau(\chi_*)} (1+z_*)^2.
\end{equation}
However, as shown in Ref.~\cite{Smith:2018bpn}, the signal-to-noise ratio of the kSZ bispectrum is dominated by the squeezed limit, as given by Eq. (2). This fact can be intuitively understood by considering that the kSZ anisotropy dominates the contributions to the CMB at $\ell\gtrsim 4000$, with a corresponding wavenumber of $\sim 1$ Mpc$^{-1}$ at relevant redshifts. Since the power in the velocity field comes mostly from large scales ($k \lesssim 0.1$ Mpc$^{-1}$), the remaining mode must have a large wavenumber in order to satisfy the triangle condition.

We note that $k_S$ falls within the nonlinear regime and make the crucial assumption that the kSZ bispectrum can be approximated by the tree-level expression, but using the nonlinear power spectrum. Ref.~\cite{Smith:2018bpn} checked the validity of this assumption against N-body simulations and found the tree-level approximation to be accurate to a few percent.

\bibliography{ref.bib}
\bibliographystyle{utcaps}
\bigskip

\end{document}